\documentclass[aps,pra,reprint,twocolumn,superscriptaddress,nofootinbib]{revtex4-1}
\pdfoutput=1 

\usepackage{graphicx}
\usepackage{bm}
\usepackage{amsmath}
\usepackage{amssymb}
\usepackage{amsfonts}
\usepackage{amsthm}
\usepackage{mathtools}
\usepackage{hyperref}
\usepackage{braket}
\usepackage[normalem]{ulem}
\usepackage{wrapfig}
\usepackage{tikz}
\usepackage{dsfont}
\usepackage{comment}

\hypersetup{colorlinks=true,urlcolor=[rgb]{0,0,0.5},citecolor=[rgb]{0.5,0,0},linkcolor=[rgb]{0,0,0.4}}

\def\and{\quad {\rm and} \quad}

\begin{document}
\title{Gravitational Constrained Instantons}

\author{Jordan Cotler}
\email{jcotler@fas.harvard.edu}
\affiliation{\it Society of Fellows, Harvard University, Cambridge, MA 02138, USA}

\author{Kristan Jensen}
\email{kristanj@uvic.ca}
\affiliation{\it Department of Physics \& Astronomy, San Francisco State University, San Francisco, CA 94132, USA}
\affiliation{\it Department of Physics and Astronomy, University of Victoria, Victoria, BC V8W 3P6, Canada}

\begin{abstract}
We find constrained instantons in Einstein gravity with and without a cosmological constant. These configurations are not saddle points of the Einstein-Hilbert action, yet they contribute to non-perturbative processes in quantum gravity.  In some cases we expect that they give the dominant contribution from spacetimes with certain fixed topologies.  With negative cosmological constant, these metrics describe wormholes connecting two asymptotic regions. We find many examples of such wormhole metrics and for certain symmetric configurations establish their perturbative stability. We expect that the Euclidean versions of these wormholes encode the energy level statistics of AdS black hole microstates. In the de Sitter and flat space settings we find new homogeneous and isotropic bounce and big bang/crunch cosmologies.
\end{abstract}


\maketitle

\textit{Introduction.}~There has been renewed interest in wormholes stemming from recent progress in low-dimensional quantum gravity.  In particular, Euclidean wormholes in Jackiw-Teitelboim (JT) and pure AdS$_3$ quantum gravity encode the energy level statistics of black hole microstates in those simple models~\cite{Saad:2018bqo, Saad:2019lba, cotler2020ads, cotler2020ads2}. However, it has been unclear if similar analyses exist in $3+1$ and higher dimensions, where pure quantum gravity requires an ultraviolet completion.  At a technical level, it is difficult to find Euclidean wormholes in $d+1\geq 3$ dimensions which are both perturbatively and non-perturbatively stable.  There is a long history of wormhole solutions, nearly all of which are now known to be unstable~\cite{giddings1988axion, maldacena2004wormholes, arkani2007euclidean, hertog2019euclidean}.  Furthermore the role of Euclidean wormholes in AdS/CFT is puzzling. If they contribute to the gravity path integral then there is some tension with the standard holographic dictionary~\cite{witten1999connectedness,maldacena2004wormholes}.

Inspired by recent progress in low-dimensional gravity~\cite{Saad:2018bqo, Saad:2019lba, cotler2019low, witten2020matrix, Stanford:2020wkf,cotler2020ads, cotler2020ads2} as well as the resolution of certain information paradoxes via replica wormholes~\cite{penington2019replica, almheiri2020replica}, we are emboldened to take Einstein gravity seriously as an effective field theory in which one not only considers solutions to the field equations, but a sum over metrics, which we attempt to make sense of as best we can.  It is not our purpose nor within our ability to address whether consistent theories of quantum gravity indeed include such a sum; rather, our goal is to mine new physics from the weakly coupled, weakly curved regimes where Einstein gravity ought to make sense as an effective field theory.

In this Letter we find new wormhole configurations in pure Einstein gravity with negative cosmological constant. These wormholes smoothly connect two asymptotic regions, and they are not solutions to Einstein's equations. Indeed, for many cases we study, there are no classical solutions of this sort in pure gravity. Rather, in the language of~\cite{affleck1981constrained} they are ``constrained instantons,'' meaning that they extremize the Einstein-Hilbert action subject to a constraint. This constraint may be understood to be the length of the wormhole, or the energy perceived by an observer on the boundary. For fixed boundary data there is a $d+1$-dimensional family of wormholes, labeled by $d$ ``twist'' moduli and a parameter which controls the size of the bottleneck of the wormhole. The Einstein-Hilbert action of the wormhole depends on the boundary data and this size parameter. 

There is good reason to expect that wormholes in Einstein gravity are constrained instantons.  Indeed, the previously studied wormholes in JT gravity and more general 2D dilaton theories~\cite{Saad:2018bqo, Saad:2019lba, cotler2019low, witten2020matrix, Stanford:2020wkf}, as well as pure AdS$_3$ gravity~\cite{cotler2020ads, cotler2020ads2}, are all examples of constrained instantons.

We find a zoo of higher-dimensional Euclidean wormholes, some of which admit a continuation to Lorentzian signature where they become traversable wormholes. These Euclidean metrics admit another continuation to new asymptotically de Sitter cosmologies.  Taking flat space limits, we find new flat space cosmologies. We showcase a few examples, and relegate others to the Appendix.  Further, we initiate a stability analysis of some of the examples. We prove that certain symmetric wormholes are perturbatively stable.  Assuming perturbative stability in general, we suggest that the one-loop approximation to the path integral over these instantons (including an integral over the instanton parameters) describes a coarse-grained approximation to the energy level statistics of AdS black hole microstates.  In this sense, Euclidean quantum gravity would provide a statistically averaged, `mesoscopic' description of microstates.

\textit{Constrained Instantons.}~An essential ingredient in our analysis is the method of constrained instantons~\cite{affleck1981constrained, frishman1979large} which we briefly review.  We begin with ordinary instantons, to contrast with the constrained case. Given a path integral description of a quantum system, a saddle is a stationary point of the classical action which can be leveraged to perform a saddle-point approximation of the path integral. An `instanton' usually refers to a localized solution, like the BPST solution of four-dimensional Yang-Mills theory, but we use it to refer to any non-trivial solution. In the $\hbar\to 0$ limit its contribution to the path integral is weighted by $e^{-S/\hbar}$ with $S$ the instanton action.

The method of constrained instantons is especially useful in situations where there are no instantons, like a Higgs phase of four-dimensional Yang-Mills theory coupled to matter.  To illustrate the idea, consider a Euclidean path integral over a field $\phi$, schematically $\int [d\phi] \exp(-S[\phi]/\hbar)$.  For instance, if the theory is a gauge theory, then $S$ includes the classical action, boundary terms, gauge fixing terms, and ghost terms.
Let $\mathcal{C}[\phi]$ be some functional of $\phi$. Then
\begin{equation}
	\int [d\phi] \,e^{-\frac{1}{\hbar}S[\phi]} = \int d\zeta \int [d\phi] \,\delta(\mathcal{C}[\phi] - \zeta) \, e^{-\frac{1}{\hbar}S[\phi]}\,,
\end{equation}
where we have introduced a constraint $\mathcal{C}[\phi]=\zeta$ but rendered it innocuous by integrating over $\zeta$. As such, the constraint need not be gauge invariant.
We can rewrite the above integral as
\begin{equation}
\label{E:Effaction1}
	\frac{1}{\hbar}\int d\lambda \int d\zeta \int [d\phi] \, e^{-\frac{1}{\hbar}\left(S[\phi] + \lambda (\mathcal{C}[\phi] - \zeta)\right)}\,,
\end{equation}
where $\lambda$ is integrated parallel to the imaginary axis.  In the full variational problem one varies with respect to $(\phi,\lambda,\zeta)$ so that the equations of motion are
\begin{equation}
\label{E:EOM1way0}
	\delta S[\phi] + \lambda\,\delta \mathcal{C}[\phi] = 0\,, \quad \mathcal{C}[\phi] = \zeta\,, \quad \lambda = 0\,,
\end{equation}
i.e.~$\delta S[\phi] = 0$. But suppose we define a new variational problem where one varies $\phi$ and $\lambda$ but keeps $\zeta$ fixed. The corresponding equations of motion are
\begin{equation}
\label{E:EOM1way1}
	\delta S[\phi] + \lambda \, \delta \mathcal{C}[\phi] = 0\,, \qquad \mathcal{C}[\phi] = \zeta\,.
\end{equation}
Any solution to the original field equation $\delta S[\phi]=0$ which satisfies the constraint solves these equations. Crucially, this variational problem may admit more solutions. We refer to solutions of~\eqref{E:EOM1way1} with $\lambda \not =0$ as `constrained instantons.'  If we have such a constrained instanton then locally in field space it is connected to a 1-parameter family of solutions $(\phi_\zeta, \lambda_\zeta)$ labeled by $\zeta$.  The reason is that perturbing $\lambda$ by a small amount acts as a source in the constrained instanton equations of motion, and these modified equations can then be solved.

Notice that there is a saddle point approximation at fixed $\zeta$\,; moreover, this is true even when the unconstrained path integral does not admit saddles (e.g.~\cite{affleck1981constrained}). This leads to a new, candidate semiclassical approximation to the full path integral: we can perform a saddle point approximation at fixed $\zeta$ to some chosen order in perturbation theory, and then integrate over $\zeta$ last. In the semiclassical limit, these saddle point contributions will clearly be important for the path integral evaluation. Let us write a formal expression for the saddle point approximation of the $(\phi_\zeta, \lambda_\zeta)$ constrained instantons.  To second order in fluctuations, the total action is
\begin{align}
\begin{split}
&S[\phi_\zeta] + \frac{1}{2} \int d^{d+1}x \, d^{d+1}y\,\delta\phi(x) \frac{\delta^2 S}{\delta \phi(x) \delta \phi(y)} \delta \phi(y) \\
& \qquad \qquad \qquad \qquad \qquad \qquad - i \,\delta \lambda \int d^{d+1}x \frac{\delta \mathcal{C}}{\delta \phi} \, \delta \phi\,.
\end{split}
\end{align}
Let $v_i$ be the bosonic eigenfunctions of $\frac{\delta^2 S}{\delta \phi^2}$ with eigenvalue $\chi_i$\,,
and denote $\int d^{d+1}x \, \frac{\delta \mathcal{C}}{\delta \phi} v_i = \kappa_i$.  Then the 1-loop approximation around the $(\phi_\zeta, \lambda_\zeta)$ constrained saddles is
\begin{equation}
\label{E:semiclassical}
\int d\zeta \, e^{-\frac{1}{\hbar} S[\phi_\zeta]} V_{\text{zm}}(\zeta) \frac{\mathcal{D}_F'(\zeta)}{\sqrt{\mathcal{D}_B'(\zeta)}} \sqrt{\frac{1}{2\pi}\sum_i \frac{\chi_i}{\kappa_i^2}}
\end{equation}
where $\mathcal{D}_B'(\zeta)$ and $\mathcal{D}_F'(\zeta)$ are the fixed bosonic and fermionic determinants (excepting zero modes) and $V_{\text{zm}}(\zeta)$ is the zero mode volume at fixed $\zeta$.

Below we set $\hbar = 1$, but keep factors of Newton's constant $G$ so that $1/G$ is our large saddle parameter.

\textit{Constrained instantons for Euclidean AdS.}~We begin with Euclidean Einstein gravity in $d+1$ dimensions and negative cosmological constant, with action
\begin{equation}
\label{E:EHaction1}
	S_{\text{EH}} =-\frac{1}{16 \pi G} \!\int_{\mathcal{M}} \!\! d^{d+1} x \sqrt{g} (R - 2 \Lambda) \,, \,\,\, \Lambda = -\frac{d(d-1)}{2L^2}\,.
\end{equation}
We impose asymptotically Euclidean AdS boundary conditions, and include appropriate boundary terms including those counterterms required by holographic renormalization~\cite{skenderis2002lecture}. Henceforth we use $L=1$ units unless noted otherwise.

The classical solution to Eqn.~\eqref{E:EHaction1} is known: it is just standard Euclidean AdS$_{d+1}$ with a single boundary.  However, our goal is to find a family of constrained instantons which correspond to Euclidean wormholes with two boundaries.  We work in a global coordinate system $(\rho, x^i)$ where $i=1,2,...,d$ and $\rho$ is a radial coordinate, and one reaches the two boundaries as $\rho \to \pm \infty$.  We fix $\rho$ so that $g_{\rho i} = 0$ and $g_{\rho \rho} = 1$.  By the logic of Eqn.~\eqref{E:Effaction1}, we need to choose an inspired constraint. In this Letter we consider two different constraints which lead to the same wormholes. The first constraint is to fix the length of the wormhole connecting the two boundaries as in~\cite{Stanford:2020wkf}. This constraint is necessarily non-covariant, and reads
\begin{equation}
\label{E:constraint1}
	\mathcal{C}[g_{\mu \nu}] = \frac{1}{8\pi G} \int d^{d+1}x \, \Lambda \, \sqrt{g_{\rho \rho}}\,F(x)\,,
\end{equation} 
for a function $F(x)$ to be chosen judiciously later.

The constraint here only depends on $g_{\rho \rho}$, and so only modifies the $\rho\rho$ component of Einstein's equation. In our radial gauge the modified Einstein's equations are
\begin{equation}
	\label{E:constrainedEinstein}
	\!\!\sqrt{g}\left( R^{\mu \nu} \! - \frac{R}{2}g^{\mu \nu} + \Lambda g^{\mu \nu}\! \right)\!+  \lambda \,F(x)\Lambda \delta^{\mu}_{\rho}\delta^{\nu}_{\rho}  = 0\,.
\end{equation}
Remarkably, we find a wealth of Euclidean wormhole solutions to this constrained problem, many of which we can write down analytically. We focus on a few simple cases, leaving a few more examples to the Appendix. 

\textbf{Torus boundary.}~Consider Euclidean wormholes connecting two regions with torus boundary. These boundaries are specified by independent conformal structures. When these structures are generic and different from each other, we expect to find a wormhole described by a line element of the form
\begin{equation}
	\label{E:AdSdgeneral}
	ds^2 = d\rho^2 + h_{ij}(\rho) dx^i dx^j \,,
\end{equation}
where the $x^i$ parameterize a $d$-dimensional torus. Although we have yet to find a wormhole connecting boundary tori with completely generic conformal structures in all dimensions (we have succeeded in three spacetime dimensions), we have found several special subclasses.  The simplest is a highly symmetric configuration which solves the modified equations of motion:\footnote{With nonzero $f^i(\rho)$ one must modify the constraint slightly so that this metric is a solution of the modified problem.}
\begin{equation}
	\label{E:AdSdsymmetric1}
	ds^2 = d\rho^2 + b^2 \left( 2\cosh\!\left(\frac{d\,\rho}{2}\right)\right)^{\frac{4}{d}} \delta_{ij} dy^i dy^j\,.
\end{equation}
Here $b> 0$ and we have relaxed our radial gauge choice slightly by allowing $y^i = x^i + f^i(\rho)$, where $f^i(\rho)$ is pure gauge unless it has support out to the boundaries. In that case it contains physical data, a relative twist $\tau^i$ between the two boundaries given by $\tau^i = \lim_{\rho\to\infty}f^i(\rho) - \lim_{\rho\to-\infty}f^i(\rho)$. This is a bottleneck geometry, with a minimal torus of volume $\sim b^d$ at $\rho=0$, so non-singularity implies $b>0$. 

\begin{figure}[t]
\includegraphics[scale=.45]{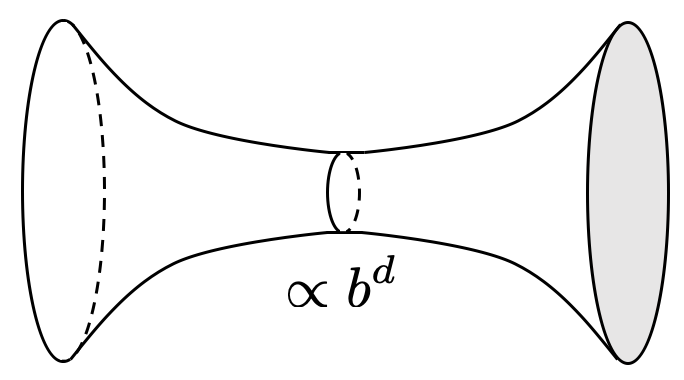}
\caption{A schematic of a wormhole with torus boundaries, and a bottleneck with volume proportional to $b^d$.}
\label{fig:wormhole1}
\end{figure}

This geometry has two boundaries as $\rho\to\pm \infty$ with aligned conformal structures specified by the same boundary metric $\delta_{ij}$. By modifying the identifications of the $x^i$, we can equip one of the boundary tori with any conformal structure we wish, but then the conformal structure of the other will be aligned with that of the first. Together, the $(b,\tau^i)$ form the instanton parameters at fixed values of the boundary conformal structures. The $\tau^i$ are large diffeomorphisms and so exact zero modes, while the renormalized action of this configuration $S_{\rm ren} = \frac{(d-1)\text{vol}(\mathbb{T}^d)}{2\pi G}\,b^d$ depends on the size $b$. Figure~\ref{fig:wormhole1} depicts the wormhole.

The metric~\eqref{E:AdSdsymmetric1} satisfies the modified equations~\eqref{E:constrainedEinstein}. It is easy to show that the metric in~\eqref{E:AdSdsymmetric1} satisfies the $ij$ and $\rho i$ components of the ordinary Einstein's equations. However, on this metric we find
\begin{equation}
	\sqrt{g}\left( R^{\rho \rho} - \frac{R}{2}g^{\rho \rho} + \Lambda g^{\rho \rho}\right) = \Lambda \,b^d \,,
\end{equation}
so that we can solve the $\rho\rho$ component with $\lambda = -b^d$ and $F(x)=1$.

In the Appendix we demonstrate two important properties of these wormholes. The first is that they are stable against quadratic fluctuations within Einstein gravity (as well as for more general low-energy effective theories of Einstein gravity coupled to matter). The second is that this is the most general wormhole where the boundary conformal structures are aligned. We show the latter by studying linearized fluctuations of~\eqref{E:AdSdsymmetric1} so that the modified line element is of the form~\eqref{E:AdSdgeneral}. Demanding that the new metric solves the modified Einstein's equations, one finds a $d(d+2)$-dimensional space of perturbations. These perturbations can be understood as changes of the $d+1$ instanton parameters $(b,\tau^i)$, a redefinition $\rho \to \rho + \varepsilon$, and perturbations in the conformal structures on the two boundaries. There are no other parameters left, which if they existed would parameterize other deformations of the wormhole consistent with the boundary conditions.

We can slightly relax the constraint that the two conformal structures are exactly aligned. A $d$-dimensional torus may be written as $\mathbb{S}^1_{\beta} \times_f \mathbb{T}^{d-1}$, where the $\mathbb{S}^1_{\beta}$ has the interpretation of Euclidean time and the $\mathbb{T}^{d-1}$ as space. Then we may find solutions where the thermal circles on each boundary have different lengths,
\begin{align}
	\label{E:AdShighertorus1}
	ds^2 &= d\rho^2 \\
	&\hspace{-.3in}+ b^2 \!\left(\!2\cosh\!\left(\frac{d\,\rho}{2}\right)\!\right)^{\frac{4}{d}}\!\bigg(\!\bigg(\frac{\beta_1 e^{\frac{d\,\rho}{2}} + \beta_2 e^{-\frac{d\,\rho}{2}}}{2 \cosh\left(\frac{d\,\rho}{2}\right)}\bigg)^2 \! (dy^{1})^2 + ds_{\mathbb{T}^{d-1}}^2 \!\!\bigg)\,, \nonumber
\end{align}
where $ds_{\mathbb{T}^{d-1}}^2 = \sum_{i=2}^d (dy^{i})^2$ and we let $x^1\sim x^1+1$. These solutions have a renormalized action
\begin{equation}
\label{E:onshell2}
	S_{\rm ren} = (\beta_1+\beta_2)E\,, \,\,\, E = \text{vol}(\mathbb{T}^{d-1})\,\varepsilon\,, \,\,\, \varepsilon = \frac{(d-1)b^d}{4\pi G}\,,
\end{equation}
with $\varepsilon$ the energy density. We will return to this observation shortly. 

For the analytic continuation $\beta_1 = i T$ and $\beta_2 = - i T$, where the action vanishes, this geometry becomes a genuine saddle for any $b$.  This particular configuration is a ``double cone'' geometry of~\cite{Saad:2018bqo}.  Accordingly, for general $\beta_1, \beta_2$, our configurations generalize the double cone.

Analytically continuing $x^1$ to real time this geometry becomes a traversable wormhole connecting the two boundaries at $\rho \to \pm \infty$: in this geometry, null geodesics at fixed location on the spatial torus take a finite time to travel from one boundary to the other.

Finally, in addition to the wormhole there is a related singular geometry, with line element
\begin{equation}
\label{E:AdScrunch}
	ds^2 = d\rho^2 + b^2 \left( 2\sinh\left(\frac{d\,\rho}{2}\right)\right)^{\frac{4}{d}}\delta_{ij}dx^idx^j\,,
\end{equation}
where the whole torus collapses at $\rho = 0$ with an ``opening angle'' $\sim b$. This example will show up later as it admits an interesting continuation to positive cosmological constant. 

\textbf{$\mathbb{S}^1 \times \mathbb{S}^{3}$ boundary.}~Let us briefly present another example where the boundaries are $\mathbb{S}^1\times \mathbb{S}^{d-1}$, which we expect to be related to black hole physics in AdS. For $d=2$ these boundaries are tori, which we covered above and is given in some more detail in the Appendix. For $d=4$ we find a simple solution with boundaries $\mathbb{S}^1_{\beta_1}\times\mathbb{S}^3$ and $\mathbb{S}^1_{\beta_2}\times\mathbb{S}^3$,
\begin{align}
\begin{split}
\label{E:AdS5bh}
	ds^2 &= d\rho^2 + \frac{b^4}{8}\frac{(\beta_1 e^{2\rho} + \beta_2 e^{-2\rho})^2}{b^2\cosh(2\rho) - 1} (dx^{1})^2  \\
	& \qquad \qquad \qquad \qquad + \frac{1}{2}\left( b^2\cosh(2\rho)-1\right) ds_{\mathbb{S}_3}^2\,,
\end{split}
\end{align} 
where $x^1\sim x^1+1$ and non-singularity implies $b>1$. There are also twist zero modes which we are neglecting to write. In a holographic renormalization scheme where one does not add a finite counterterm $\propto \int d^4x \sqrt{\gamma}\,R_{\gamma}^2$, the renormalized action is
\begin{equation}
\label{E:onshell3}
	S_{\rm ren} =  (\beta_1+\beta_2)E\,, \,\,\, E =b^4 E_0\,, \,\,\, E_0 = \frac{3}{8}\frac{\text{vol}(\mathbb{S}^3)}{16\pi G}\,,
\end{equation}
with $E_0$ the energy threshold of small black holes in AdS$_5$ in this scheme. 

As with the torus wormholes, these metrics describe traversable wormholes upon continuing to real time. Further for $\beta_1 = - \beta_2 = i T$ these metrics are genuine saddles for all $b$, and reduce to a double cone geometry of~\cite{Saad:2018bqo}.

\textit{Action and boundary stress tensor.}~The constrained instantons in the last Section have the property that, despite being off-shell configurations, their action evaluates to a pure boundary term.  In the Appendix we show that this follows from the $ij$ components of the modified Einstein's equations~\eqref{E:constrainedEinstein}, as well as derive a simple expression for the action of those wormholes in terms of the holographic stress tensor $T^{ij}$~\cite{balasubramanian1999stress}. Working in a gauge where our metrics take the form $ds^2 = d\rho^2 + h_{ij}(x,\rho) dx^i dx^j$, the renormalized action for wormholes with torus or $\mathbb{S}^1\times \mathbb{S}^{3}$ boundary is
\begin{equation}
	S_{\rm ren} =\frac{1}{d}\int_{\partial \mathcal{M}} \!\! d^{d}x \sqrt{\gamma} \, \gamma_{ij} T^{ij}\,,
\end{equation}
where $\gamma$ is the boundary metric and the action is the sum of two terms, one for each component of the boundary.

We infer that the boundary stress tensor has nonzero trace for these wormholes. This is in contrast with on-shell configurations of Einstein gravity, where the trace of the boundary stress tensor is fixed in terms of the boundary metric~\cite{Henningson:1998gx}. For the torus wormhole in~\eqref{E:AdShighertorus1} the energy density perceived on the two boundaries is identical and is given by
\begin{equation}
	\varepsilon_1 = \varepsilon_2 = \frac{(d-1)b^d}{4\pi G}\,, 
\end{equation}
while the stress tensor trace on boundary 1 is
\begin{equation}
	\frac{1}{d}(\gamma_{ij}T^{ij})_1 = \frac{(d-1)b^d}{8\pi G}\left( 1 + \frac{\beta_2}{\beta_1}\right)\,,
\end{equation}
with a similar expression on boundary 2.

So far we have found constrained instantons by fixing the length between the two boundaries. We would like another, covariant constraint that leads to the same wormholes. Given the boundary stress tensors here, instead we can fix the energies $E_1, E_2$ perceived on the two boundaries as in~\cite{Saad:2018bqo,Stanford:2020wkf}. Indeed our wormholes extremize the Einstein-Hilbert action subject to the constraint that $E_1=E_2=E$, as long as $E$ is above a critical threshold (e.g.~the small black hole threshold for the $\mathbb{S}^1\times\mathbb{S}^3$ case). This likely implies that the wormholes studied in this Letter give the dominant contribution to the gravity path integral from spacetimes with this simple topology since we expect that more general wormholes are labeled by fixed $E$, and at least within the subset of metrics of a fixed $E$, our wormholes extremize the action.

\textit{de Sitter and flat space.}~So far we have focused on configurations in gravity with negative cosmological constant.  It is natural to ask if there are similar constrained instantons with positive or zero cosmological constant.

Consider a positive cosmological constant. There is a simple procedure to go from our Euclidean AdS constrained instantons to de Sitter versions thereof:\footnote{There is another continuation introduced in~\cite{Maldacena:2010un} which can be adapted here, amounting to a shift $\rho \to t - i a$ for some $a$. That continuation has not led us to any new geometries.}  (i) restore the AdS radius $L$, (ii) send $L \to i L$, and (iii) Wick rotate $\rho = it$. (Accordingly $\rho/L \to t/L$.)  For example, starting with the symmetric wormhole in Eqn.~\eqref{E:AdSdsymmetric1}, we obtain
\begin{equation}
\label{E:dSsymmetric1}
	ds^2 = - dt^2 + b^2\left( 2 \cosh\left(\frac{d\, t}{2L}\right)\right)^{\frac{4}{d}} \delta_{ij} dy^i dy^j\,,
\end{equation}
where $y^i = x^i + f^i(\rho)$ as usual. This is a bounce cosmology with a flat, toroidal universe. In our study of wormholes at negative cosmological constant, the constraint modified the $\rho\rho$ component of Einstein's equations. In this case the constraint, effectively the time elapsed between past and future infinity, modifies the $tt$ component. So from the point of view of an observer in this spacetime who assumes that the usual Einstein's equations are satisfied, that observer would infer that the spacetime is supported by a negative energy density (in addition to the cosmological constant)
\begin{equation}
	T^{tt}_{\rm eff} = -\frac{1}{\sqrt{-g}}\frac{d(d-1)}{4\pi GL^2}\,b^d\,.
\end{equation}
We stress that this apparent energy density is fictitious. In our approach the spacetime~\eqref{E:dSsymmetric1} is not sourced by any matter; it is a constrained instanton of Einstein gravity itself, and so is not a solution to the usual Einstein's equations. We also stress that this cosmological process is classically forbidden in Einstein gravity without matter; it is intrinsically quantum mechanical.

We construct other asymptotically de Sitter bounce geometries in the Appendix with closed and open universes. 

A rather dramatic class of de Sitter constrained instantons correspond to big bang/crunch cosmologies. Starting with Eqn.~\eqref{E:AdScrunch} we find
\begin{equation}
\label{E:dSbigbang1}
	ds^2 = - dt^2 + b^2\left( 2 \sinh\left(\frac{d  |t|}{2L}\right) \right)^{\frac{4}{d}}\delta_{ij} dx^i dx^j\,,
\end{equation}
for either $t \geq 0$ or $t \leq 0$, corresponding to a big bang or big crunch respectively, with torus boundary.  An observer in this universe might infer that the spacetime is supported by a positive energy density
\begin{equation}
	T^{tt}_{\rm eff} = \frac{1}{\sqrt{-g}}\frac{d(d-1)}{4\pi GL^2} \, b^d\,.
\end{equation}
In the Appendix we also find big bang/crunch cosmologies with a closed or open universe.  For example a simple $d=2$ instanton with a closed universe (which is classically forbidden) is given by
\begin{equation}
	ds^2 = - dt^2 + 4b^2 \sinh^2\left(\frac{t}{L}\right) ds_{\mathbb{S}_2}^2\,.
\end{equation}
Other explicit examples are given in the Appendix. All big bang/crunch cosmologies we find are supported by a fictitious positive energy density as above. We do not have explicit solutions in $3+1$ dimensions in all cases, but can solve for the warp factor numerically.

To summarize the de Sitter solutions, there are novel, natural cosmologies (including big bangs/crunches) which are not classically allowed, yet contribute to the gravity path integral. Such solutions may have the potential for real-world physical significance. For instance, they may offer interesting alternatives to the Hartle-Hawking state~\cite{hartle1983wave}. Or more dramatically, perhaps we live in a constrained instanton cosmology, with parameters like $b$ accounting for part of the observed energy budget of the universe.

To obtain flat space constrained instantons from either the AdS or dS solutions, we simply restore $L$ and take $L \to 0$.  To obtain non-trivial solutions, sometimes we will need to make parameters, such as the bottleneck scale $b$, scale with $L$ as well.  For instance, the flat space limit of Eqn.~\eqref{E:dSsymmetric1} is just ordinary flat space $\mathbb{R} \times \mathbb{T}^{d-1}$; the flat space limit of Eqn.~\eqref{E:dSbigbang1} with $b^2 = \left( \frac{L}{d} \right)^{\frac{4}{d}} B^2$ is
\begin{equation}
ds^2 = -dt^2 + B^2 |t|^{\frac{4}{d}} \delta_{ij} dx^i dx^j
\end{equation}
which is a big bang/crunch cosmology. Other novel solutions are collected in the Appendix.

\textit{Statistics of black hole microstates.}~So far we have found families of constrained instantons labeled by the instanton parameters $(b,\tau^i)$. To compute the semiclassical approximation to the path integral as in~\eqref{E:semiclassical} we require constrained saddles as well as the appropriate one-loop determinants around them. Those determinants may not be tractable for Einstein gravity in more than three spacetime dimensions.  

What does the wormhole amplitude encode? In JT gravity and pure AdS$_3$ gravity the full wormhole amplitude has been computed in~\cite{Saad:2019lba, cotler2020ads, cotler2020ads2}. (In three dimensions the amplitude was computed using a constrain first approach in the first-order formalism~\cite{cotler2020ads}, rather than by performing a sum over metrics; the amplitude was also computed via a bootstrap method in~\cite{cotler2020ads2}.) In those examples this basic two-boundary wormhole encodes the leading two-point energy fluctuation statistics of black hole microstates, and in particular those statistics match a random matrix theory prediction.

In more than three spacetime dimensions there is much less that we can say without knowing the integration measure over $b$. However, we would like to point out a route by which the one-loop approximation to the wormhole amplitude may yield the same random matrix theory answer as in the lower dimensions, at least in the low-temperature limit. The torus and $\mathbb{S}^1\times \mathbb{S}^3$ wormhole actions take the simple form $(\beta_1+\beta_2)E$ where $E=V\varepsilon \propto \frac{V b^d}{G}$ with $V$ the spatial volume. (These wormholes also carry zero angular momentum, and so would encode the spectral statistics of spinless primaries.) Suppose the quantum-corrected measure over $b$ (after integrating out all other fluctuations) is just $d\!\left( \frac{b^d}{G}\right)\sim d\varepsilon$. The integration over the twist zero modes will produce a factor $\mathcal{V}_0 = \sqrt{\beta_1\beta_2}\,V$ (see~\cite{cotler2020ads} for a discussion in the context of 3D gravity). We would then have a one-loop amplitude
\begin{equation}
	\sim \mathcal{V}_0 \int d\varepsilon\, e^{-(\beta_1+\beta_2)V \varepsilon} = \frac{\sqrt{\beta_1\beta_2}}{\beta_1+\beta_2} \,e^{-(\beta_1+\beta_2)E_0}\,,
\end{equation}
where $E_0$ is the energy at the lower bound of integration. For a normalization constant $\frac{1}{2\pi}$ this expression would match a random matrix theory prediction, where $E_0$ corresponds to the spinless black hole threshold. This random matrix theory prediction is essentially the physics of level repulsion, which is a generic feature of many-body chaotic quantum systems upon coarse-graining. We cannot help but note that the torus and $\mathbb{S}^1\times \mathbb{S}^3$ wormholes carry energies $E\geq  E_0$. Clearly much more work remains to be done.

\textit{Discussion.}~We have established the existence of constrained instantons in Einstein gravity with and without cosmological constant. With negative cosmological constant, relevant for the AdS/CFT correspondence, we found explicit solutions corresponding to Euclidean wormholes with a variety of boundary topologies. Further, they may give the dominant contribution from two-sided wormholes with trivial bulk topology in pure quantum gravity.

In two and three spacetime dimensions~\cite{Cotler:2016fpe, Saad:2018bqo, Saad:2019lba, cotler2020ads, cotler2020ads2} the statistical properties of AdS black hole microstates are encoded in smooth geometries, albeit constrained instantons (Euclidean wormholes) rather than solutions to the field equations. In JT gravity and pure AdS$_3$ gravity, the two-point fluctuation statistics computed from the simplest wormholes are exactly what one anticipates from a random matrix theory description. It is then striking that there are similar Euclidean wormholes in pure Einstein gravity in $3+1$ and higher spacetime dimensions. With the lower-dimensional results and the universality of random matrix theory in mind, it is tempting to speculate that the path integral over these constrained instantons gives a coarse-grained approximation to the level statistics of AdS black hole microstates. As such, the gravitational path integral would be a \textit{mesoscopic} description of quantum gravity, which does not know about the precise spectrum of black hole microstates (which in tractable examples depends on the UV completion), but instead provides statistical information.

Recent works have advocated for an \textit{ensemble-averaged} description of JT gravity and pure gravity in three spacetime dimensions~\cite{Cotler:2016fpe, Saad:2018bqo, Saad:2019lba, cotler2020ads, cotler2020ads2, belin2020random, 1800406, 1800422, maxfield2020path}, i.e.~dualities between pure quantum gravity and a disordered system. JT gravity is renormalizable, and pure AdS$_3$ gravity is power-counting renormalizable, and so modulo the convergence of the sum over topologies, these models do not require a UV completion. In higher spacetime dimension, pure gravity is emphatically non-renormalizable, but amazingly our low-energy analysis may yield sensible answers for certain statistical quantities probing gravitational microstates. We expect that these wormhole amplitudes are akin to the Bekenstein-Hawking entropy, in that they use the effective IR description of quantum gravity to approximate very particular properties of black holes (like the density of states or fluctuation statistics) which na\"{i}vely require UV data to reliably compute.

The wormhole metrics that we found are in general macroscopic and smooth, and are within the purview of effective field theory.  We caution, however, that curvatures blow up when the bottleneck is small ($b \to 0$ for the torus wormholes and $b \to 1$ for the $\mathbb{S}^1 \times \mathbb{S}^3$ wormholes).  Nonetheless, outside of this regime, the sum over the wormhole metrics gives a controlled contribution to the gravitational path integral.

We have found that the symmetric AdS$_{d+1}$ wormholes are perturbatively stable within Einstein gravity, which poses a potential conflict with paradigmic examples of the AdS/CFT correspondence, where certain string theories are equated with single instances of a CFT, rather than an ensemble. With this tension in mind, in an upcoming work~\cite{CJWIP2} we embed the wormholes studied here into string theory in various settings, and perform a stability analysis within perturbative string theory. We also study the prospect of stabilizing these wormholes with suitable boundary conditions for bulk matter as in~\cite{maldacena2018eternal}.

In our previous studies of de Sitter JT gravity~\cite{cotler2019low, cotler2019emergent} we have seen how constrained instantons in de Sitter encode the transition amplitudes of classically forbidden processes, like the nucleation or annihilation of a closed universe. The higher-dimensional cosmologies introduced in this Letter may provide a starting point for similar studies in Einstein gravity, along with new quantum states of the universe generalizing the no-boundary proposal. 

There are many future directions suggested by our findings. One is the development of numerical methods for finding constrained instantons, which should be able to leverage existing techniques from numerical relativity.  Relatedly, it seems plausible that there exist gravitational constrained instantons with more than two asymptotic boundaries, and perhaps numerical methods may aid in finding these solutions.  Another direction is to find constrained instantons for gravity plus matter fields or gauge fields, or even to explore constrained instantons in pure supergravity. Finally, we raised the prospect that perhaps we live in a constrained instanton cosmology. Is this so? It may be useful to find experimental constraints on the instanton parameters, which support the instanton spacetime in a way that mimics cold dark matter. 

\emph{ Acknowledgments.}
We would like to thank D.~Stanford for enlightening discussions. JC is supported by a Junior Fellowship from the Harvard Society of Fellows. KJ is supported in part by the US Department of Energy under grant number DE-SC 0013682.

\bibliographystyle{apsrev4-1}
\bibliography{refs}

\pagebreak
\appendix
\onecolumngrid

\section{Zoology of constrained instanton solutions}
In this Appendix, we enumerate the constrained instantons we have found, emphasizing simple, closed-form expressions.  Often we will restrict the dimension in order to provide non-numerical expressions for the warp factors.  For AdS and dS, we explicitly write the AdS/dS radius $L$.  In Euclidean signature, our coordinates will be $(\rho, x^i)$ for $i=1,2,...,d$, and in Lorentzian signature we use $(t,x^i)$ for $i=1,2,...,d$.  We will not discuss twists, i.e.~using coordinates $y^i = x^i + f^i(\rho)$, although they can be included.  In the asymptotically flat space setting, we do not include $L \to \infty$ degenerations of the AdS/dS solutions that yield ordinary instantons (i.e., we only include constrained instantons).

In the context of open boundaries, namely~$\mathbb{H}^d$, our results hold if we further consider quotients by a subset of the isometry group which acts freely.

\subsection*{Asymptotically Euclidean Anti-de Sitter}
\noindent \textbf{Flat spatial slices.} \\ \\
\noindent \textit{$\mathbb{T}^d$ boundaries, non-singular solutions.}  In arbitrary $d$, we find Euclidean wormholes solutions  where the boundaries have topologies $\mathbb{S}_{\beta_1}^1 \times \mathbb{T}^{d-1}$ and $\mathbb{S}_{\beta_2}^1 \times \mathbb{T}^{d-1}$, where $\mathbb{S}_\beta$ denotes the non-topological data that the corresponding circle has circumference $\beta$. The explicit solutions are
\begin{equation}
\label{E:AppAdSTorus}
	ds^2 = d\rho^2  + b^2 \left( 2\cosh\left(\frac{d\,\rho}{2L}\right)\right)^{\frac{4}{d}}\!\left(\!\left(\frac{\beta_1 e^{\frac{d\,\rho}{2L}} + \beta_2 e^{-\frac{d\,\rho}{2L}}}{2 \cosh\left(\frac{d\,\rho}{2L}\right)}\right)^2 \! (dx^1)^2 + ds_{\mathbb{T}^{d-1}}^2 \!\!\right)
\end{equation}
where $ds_{\mathbb{T}^{d-1}}^2 = \sum_{i=1}^d (dx^i)^2$ and $b \in (0,\infty)$.  A particularly symmetric set of solutions can be found for $\beta_1 = \beta_2 $, which analytically continue (via Wick rotation) to traversable wormholes.

For $d = 2$, Eqn.~\eqref{E:AppAdSTorus} in fact represents the most general wormhole with topology $\mathbb{T}^2 \times I$, after accounting for the freedom in the identifications of the $x^i$.  We will have more to say about the $d = 2$ case in the ``Other boundary topologies and miscellany'' subsection below.
\\ \\
\textit{$\mathbb{T}^d$ boundaries, singular solutions.} These solutions are singular counterparts to Eqn.~\eqref{E:AppAdSTorus} above, 
\begin{equation}
\label{E:AppAdSTorus2}
	ds^2 = d\rho^2  + b^2 \left( 2\sinh\left(\frac{d\,\rho}{2L}\right)\right)^{\frac{4}{d}}\!\left(\!\left(\frac{\beta_1 e^{\frac{d\,\rho}{2L}} + \beta_2 e^{-\frac{d\,\rho}{2L}}}{2 \sinh\left(\frac{d\,\rho}{2L}\right)}\right)^2 \! (dx^1)^2 + ds_{\mathbb{T}^{d-1}}^2 \!\!\right)\,.
\end{equation}
These solutions seem most suitable for either $\rho \leq 0$ and $\rho \geq 0$.
\\ \\ 
\noindent \textbf{Closed spatial slices.}
\\ \\
\textit{$\mathbb{S}^d$ boundaries, non-singular solutions.}  We have a general solution of the form
\begin{equation}
	ds^2 = \frac{f(\rho/L)^{\frac{d}{2} - 2} f'(\rho/L)^2}{C + 4 f(\rho/L)^{\frac{d}{2}-1} + 4 f(\rho/L)^{\frac{d}{2}}}\, d\rho^2 + L^2 f(\rho/L) \, ds_{\mathbb{S}^d}^2\,,
\end{equation}
where $C$ is a constant.  To obtain non-singular solutions with the appropriate asymptotic behavior, we consider the case of $d$ even and let
\begin{equation}
	C = - 4 b^{d-2} (1+b^2) \,, \qquad f(\rho) = b^2 (\rho^2 + 1)\,,
\end{equation}
where $b >0$.

In some dimensions, the warp factors can be conveniently written in terms of hypertrigonometric functions.  For example, in $d = 2$ and $d = 4$ we respectively have
\begin{align}
\label{E:AdSspherehyper2}
	d = 2: \qquad ds^2 &= d\rho^2 +  L^2 b^2 \cosh^2\left(\frac{\rho}{L}\right) \,ds_{\mathbb{S}^2}^2\,, 
	\\
\label{E:AdSspherehyper4}
	d = 4: \qquad ds^2 &= d\rho^2 + \frac{L^2}{2} \left(b^2 \cosh\left(\frac{2\rho}{L}\right) -1\right)  \,ds_{\mathbb{S}^4}^2 \,,
\end{align}
where $b>0$ in $d=2$ and $b>1$ in $d=4$.
\\ \\
\textit{$\mathbb{S}^d$ boundaries, singular solutions.}  We have explicit solutions by looking at singular counterparts of Eqn.'s~\eqref{E:AdSspherehyper2} and~\eqref{E:AdSspherehyper4}, namely
\begin{align}
\label{E:AdSspherehyper2sing}
	d = 2: \qquad ds^2 &= d\rho^2 +  L^2 b^2 \sinh^2\left(\frac{\rho}{L}\right) \,ds_{\mathbb{S}^2}^2\,, 
	\\
\label{E:AdSspherehyper4sing}
	d = 4: \qquad ds^2 &= d\rho^2 + L^2 \sinh\left(\frac{\rho}{L}\right)\left(b^2 \cosh\left(\frac{\rho}{L}\right) + \sinh\left(\frac{\rho}{L}\right) \right) \,ds_{\mathbb{S}^4}^2 \,,
\end{align}
for $b >0$.
\\ \\ 
\noindent \textbf{Open spatial slices.} \\ \\
\textit{$\mathbb{H}^d$ boundaries, non-singular solutions.}  Letting $ds_{\mathbb{H}^d}^2 = \frac{1}{(x^1)^2} \sum_{i=1}^d (dx^i)^2$, we find the explicit solutions in $d = 2$ and $d = 4$:
\begin{align}
\label{E:AdShyp2}
	d = 2: \qquad ds^2 &= d\rho^2 +  b^2 \cosh^2\left(\frac{\rho}{L}\right) \,ds_{\mathbb{H}^2}^2\,, 
	\\
	\label{E:AdShyp4}
	d = 4: \qquad ds^2 &= d\rho^2 + \frac{L^2}{2} \left( b^2 \cosh\left(\frac{2\rho}{L}\right) +1\right) \,ds_{\mathbb{H}^4}^2 \,,
\end{align}
where here $b>0$.  We note that the above are genuine solutions to Einstein's equations for $b = 1$~\cite{maldacena2004wormholes}.
\\ \\
\textit{$\mathbb{H}^d$ boundaries, singular solutions.} We find explicit solutions which are singular counterparts to Eqn.'s~\eqref{E:AdShyp2} and~\eqref{E:AdShyp4}, namely
\begin{align}
\label{E:AdShyp2sing}
	d = 2: \qquad ds^2 &= d\rho^2 +  b^2 \sinh^2\left(\frac{\rho}{L}\right) \,ds_{\mathbb{H}^2}^2\,, 
	\\
	\label{E:AdShyp4sing}
	d = 4: \qquad ds^2 &= d\rho^2 + L^2 \sinh\left(\frac{\rho}{L}\right)\left( b^2 \cosh\left(\frac{\rho}{L}\right) - \sinh\left(\frac{\rho}{L}\right)\right) ds_{\mathbb{H}^4}^2 \,,
\end{align}
with $b >0$ as usual.
\\ \\ 
\textbf{Other boundary topologies and miscellany.} \\ \\
\textit{$\mathbb{S}^1 \times \mathbb{S}^{d-1}$ boundaries, non-singular solutions.} For $d = 2$, we have $\mathbb{S}^1 \times \mathbb{S}^{1} \simeq \mathbb{T}^2$ which reduces to the torus case studied above.  For higher $d$, we find explicit warp factors for $d = 4$ with boundaries $\mathbb{S}_{\beta_1}^1 \times \mathbb{S}^3$ and $\mathbb{S}_{\beta_2}^1 \times \mathbb{S}^3$, namely
\begin{equation}
ds^2 = d\rho^2 +\frac{L^2 b^4}{8}\frac{(\beta_1 e^{2\rho/L} + \beta_2 e^{-2\rho/L})^2}{b^2\cosh\left(\frac{2\rho}{L}\right) - 1} \,dx_1^2 + \frac{L^2}{2}\left( b^2\cosh\left(\frac{2\rho}{L}\right)-1\right) ds_{\mathbb{S}^3}^2\,,
\end{equation}
where $b >1$.  This can be expressed in other coordinates as
\begin{equation}
ds^2 = \frac{d\rho^2}{(\rho/L)^2 + 2}  + \frac{L^2 b^4}{8}\frac{((\rho/L)^2 + 1)^2}{b^2((\rho/L)^2 + 1) - 1} \, dx_1^2 +\frac{L^2}{2} \left(b^2((\rho/L)^2 + 1) -1\right) \, ds_{\mathbb{S}^3}^2\,.
\end{equation}
\\ \\
\textit{AdS$_3$ wormholes with $\mathbb{T}^2$ boundaries.} Consider Euclidean wormholes connecting two torus boundaries in three dimensions. In this case we can explicitly write down the most general wormhole, allowing the two boundaries to have arbitrary complex structures.  Let $A$ and $B$ be real-valued $2 \times 2$ commuting matrices. Then
\begin{equation}
\label{E:AdS3wormholes1}
	ds^2 = d\rho^2 + d\textbf{y}^T \cdot (A^T e^{\rho} + B^T e^{-\rho}) (A e^{\rho} + B e^{-\rho}) \cdot d\textbf{y}\,
\end{equation}
where $y^i = x^i + f^i(\rho)$ is the most general constrained instanton corresponding to a connected wormhole. With $x^1\sim x^1+1$ and $x^2\sim x^2+1$ it has a renormalized action
\begin{equation}
\label{E:onshell1}
	S_{\rm ren} = \frac{1}{4 \pi G}\,\text{tr}\!\left(\text{adj}\!\left(\sqrt{A^T A}\right) \sqrt{B^T B}\right)
\end{equation}
where $\text{adj}(M) = \text{det}(M)M^{-1}$ denotes the adjugate matrix (and so the above is symmetric in $A$ and $B$).  

To see that this is equivalent to our earlier result~\eqref{E:AppAdSTorus}, let us go to a basis which simultaneously diagonalizes $A$ and $B$. In that basis (which we still label as $y^i$) we have
\begin{equation}
	ds^2 = d\rho^2 + \left(\lambda^{(A)}_1 e^{\rho}+\lambda^{(B)}_1 e^{-\rho}\right)^2 (dy^1)^2 + \left(\lambda^{(A)}_2 e^{\rho}+\lambda^{(B)}_2 e^{-\rho} \right)^2 (dy^2)^2\,.
\end{equation}
Then upon a translation $\rho \to \rho + \text{constant}$ and a rescaling $y^2 \to \alpha y^2$ we can arrange for the warp factor in front of $(dy^2)^2$ to be $4b^2 \cosh^2(\rho)$. The parameters $\lambda^{(A)}_1$ and $\lambda^{(B)}_1$ can then be mapped to the parameters $\beta_1$ and $\beta_2$.

Let us briefly count the number of available parameters here, and map it to the wormhole moduli space. From the point of view of~\eqref{E:AppAdSTorus} the geometry is specified by the size parameter $b$, the twists $\tau^i$, as well as $(\beta_1,\beta_2)$ and the identifications of the $x^i$. Those identifications can be thought of as parameterizing two vectors in the complex plane, na\"{i}vely four parameters, however we are overcounting. A simultaneous rotation of these vectors produces the same torus, and a simultaneous rescaling is equivalent to a change of $b$. So the new data in these two vectors is simply the complex structure of the torus thus parameterized. So all in all the parameters here are $b$, the twists $\tau^i$, $\beta_1$ and $\beta_2$, and the complex structure of the torus parameterized by the $x^i$. These seven real parameters exactly map onto the complex structures $\tau_1$ and $\tau_2$ of the boundary tori and the three wormhole parameters $(b,\tau^i)$.
\\ \\
\textit{Linearized perturbations of symmetric wormholes with $\mathbb{T}^d$ boundaries.}
Let us examine the wormholes in Eqn.~\eqref{E:AppAdSTorus} for $\beta_1 = \beta_2 = 1$, which we call the ``symmetric wormholes'' since the boundary tori have the same conformal structure.  The most general $O(\epsilon)$ perturbation which is still a constrained instanton is
\begin{equation}
	ds^2 = d\rho^2 +b^2 \left(2\cosh\left(\frac{d\,\rho}{2 L}\right)\right)^{\frac{4}{d}} \left[\delta_{ij} + \epsilon\left(C_{ij} + D_{ij} \, \tanh\left(\frac{d\,\rho}{2L}\right) \right)\right]dy^i dy^j\,,
\end{equation}
where $C$ and $D$ are $d \times d$ symmetric matrices, and so it seems that, besides the twists, linearized perturbations are labeled by $d(d+1)$ parameters. The trace of $C$ can be mapped to a perturbation of $b$ and the trace of $D$ can be removed by a radial translation $\rho\to\rho + \text{constant}$. So besides the wormhole parameters $(b,\tau^i)$ there are $d(d+1)-2=2\left( \frac{d(d+1)}{2}-1\right)$ parameters left labeling the perturbation. But this is precisely the right number to account for perturbations of the two boundary conformal structures, which are encoded in boundary metrics (hence the $\frac{d(d+1)}{2}$) modulo independent Weyl rescalings (hence the $-1$). This counting exercise shows that, with fixed boundary conditions, the wormhole parameters are $(b,\tau^i)$, no more and no less.
\\ \\
\textit{Alternative coordinates for symmetric wormholes with $\mathbb{T}^d$ boundary.} Consider again the wormholes in Eqn.~\eqref{E:AppAdSTorus} with $\beta_1 = \beta_2 = 1$.  In even $d$, the metric satisfies the constrained instanton equations if
\begin{equation}
	ds^2 = \frac{f(\rho/L)^{\frac{d}{2} - 2} f'(\rho/L)^2}{C + 4 f_2(\rho/L)^{\frac{d}{2}}} \, d\rho^2 + f_2(\rho/L) \, ds_{\mathbb{T}^d}^2\,.
\end{equation}
and $C$ is a constant.  We can let $f(\rho) = b^2(\rho^2 + 1)$ and $C = - 4 b^d$ to get a (non-singular) wormhole metric with appropriate asymptotic behavior.  Some  examples are:
\begin{align}
	d = 2 : \qquad & \frac{1}{(\rho/L)^2 +1} \, d\rho^2 + b^2((\rho/L)^2 + 1) ds_{\mathbb{T}^2}^2 \,,
	\\
	d = 4 \, : \qquad & \frac{1}{(\rho/L)^2 +2} \, d\rho^2 + b^2((\rho/L)^2 + 1) ds_{\mathbb{T}^4}^2\,,
\end{align}
where as usual, $b>0$.

\subsection*{Asymptotically de Sitter}

Here we treat the case of asymptotically de Sitter spacetimes; many of the solutions here are counterparts of the Euclidean Anti-de Sitter spacetimes written above. \\ \\
\noindent \textbf{Flat spatial slices.} \\ \\
\noindent \textit{$\mathbb{T}^d$ boundaries, non-singular solutions.}  These solutions are related to Eqn.~\eqref{E:AppAdSTorus} above by taking $L \to iL$, $\rho = it$.  We find the following solutions with asymptotic boundaries $\mathbb{S}_{\beta_1}^1 \times \mathbb{T}^{d-1}$ and $\mathbb{S}_{\beta_2}^1 \times \mathbb{T}^{d-1}$\,:
\begin{equation}
\label{E:AppdSTorus}
	ds^2 = -dt^2  + b^2 \left( 2\cosh\left(\frac{dt}{2L}\right)\right)^{\frac{4}{d}}\!\left(\!\left(\frac{\beta_1 e^{\frac{d t}{2L}} + \beta_2 e^{-\frac{d t}{2L}}}{2 \cosh\left(\frac{dt}{2L}\right)}\right)^2 \! d(y^1)^2 + ds_{\mathbb{T}^{d-1}}^2 \!\!\right)\,,
\end{equation}
for $b >0$, which are analogs of global dS$_{d+1}$ but with torus boundaries.
\\ \\
\textit{$\mathbb{T}^d$ boundaries, singular solutions.}  The singular counterparts to Eqn.~\eqref{E:AppdSTorus} are
\begin{equation}
\label{E:AppdSTorusBang}
ds^2 = -dt^2  + b^2 \left( 2\sinh\left(\frac{d|t|}{2L}\right)\right)^{\frac{4}{d}}\!\left(\!\left(\frac{\beta_1 e^{\frac{dt}{2L}} + \beta_2 e^{-\frac{d t}{2L}}}{2 \sinh\left(\frac{dt}{2L}\right)}\right)^2 \! d(x^1)^2 + ds_{\mathbb{T}^{d-1}}^2 \!\!\right)
\end{equation}
for $b \in (0,\infty)$, which are big bang cosmologies for $t \in [0,\infty)$ and big crunch cosmologies for $t \in (-\infty,0]$.
\\ \\ 
\noindent \textbf{Closed spatial slices.} \\ \\
\textit{$\mathbb{S}^d$ boundaries, non-singular solutions.} We find explicit solutions in $d = 2$ and $d = 4$, which are de Sitter analogs of Eqn.'s~\eqref{E:AdSspherehyper2} and~\eqref{E:AdSspherehyper4}.  In particular, we have
\begin{align}
\label{E:dSspherehyper2}
d = 2: \qquad ds^2 &= -dt^2 +  L^2 b^2 \cosh^2\left(\frac{t}{L}\right) \,ds_{\mathbb{S}^2}^2 \,,\\
\label{E:dSspherehyper4}
d = 4: \qquad ds^2 &= -dt^2 + \frac{L^2}{2} \left(b^2\cosh\left(\frac{2t}{L}\right) + 1\right)  ds_{\mathbb{S}^4}^2 \,.
\end{align}
When $b=1$ these metrics are genuine saddles, giving global de Sitter space, while for other $b >0$ we have constrained instanton versions of global de Sitter.
\\ \\
\textit{$\mathbb{S}^d$ boundaries, singular solutions.} The singular analogs of Eqn.'s~\eqref{E:dSspherehyper2} and~\eqref{E:dSspherehyper4} are
\begin{align}
\label{E:dSspherehyper2Bang}
	d = 2: \qquad ds^2 &= -dt^2 +  L^2 b^2 \sinh^2\left(\frac{t}{L}\right) \,ds_{\mathbb{S}^2}^2 \,,
	\\
	\label{E:dSspherehyper4Bang}
	d = 4: \qquad ds^2 &= -dt^2 + L^2 \sinh\left(\frac{t}{L}\right)\left(b^2\cosh\left(\frac{t}{L}\right) - \sinh\left(\frac{t}{L}\right)\right)  ds_{\mathbb{S}^4}^2 \,,
\end{align}
for $b >0$, which are big bang/crunch cosmologies.
\\ \\ 
\noindent \textbf{Open spatial slices.} \\ \\
\textit{$\mathbb{H}^d$ boundaries, non-singular solutions.}  The de Sitter counterpart to Eqn.'s~\eqref{E:AdShyp2} and~\eqref{E:AdShyp4} in $d = 2$ and $d = 4$, respectively, are
\begin{align}
\label{E:dShyp2}
	d = 2: \qquad ds^2 &= -dt^2 +  b^2 \cosh^2\left(\frac{t}{L}\right) \,ds_{\mathbb{H}^2}^2 \,,
	\\
	\label{E:dShyp4}
	d = 4: \qquad ds^2 &= -dt^2 + \frac{L^2}{2} \left(b^2\cosh\left(\frac{2t}{L}\right) +1\right) \,ds_{\mathbb{H}^4}^2 \,,
\end{align}
for $b >0$.
\\ \\
\textit{$\mathbb{H}^d$ boundaries, singular solutions.}  The singular, cosmological analogs of Eqn.'s~\eqref{E:dShyp2} and~\eqref{E:dShyp4} immediately above are
\begin{align}
	\label{E:dShyp2}
	d = 2: \qquad ds^2 &= -dt^2 +  b^2 \sinh^2\left(\frac{t}{L}\right) \,ds_{\mathbb{H}^2}^2\,, 
	\\
	\label{E:dShyp4}
	d = 4: \qquad ds^2 &= -dt^2 + L^2 \sinh\left(\frac{t}{L}\right) \left(b^2 \cosh\left(\frac{t}{L}\right) + \sinh\left(\frac{t}{L}\right) \right)\,ds_{\mathbb{H}^4}^2 \,,
\end{align}
again for $b >0$.

\subsection*{Asymptotically flat}

The solutions in here can be obtained by appropriate flat space limits of the previously described AdS and dS constrained instantons.  We omit the case of non-singular solutions for $\mathbb{T}^d$ boundaries, since here we just find flat space on $\mathbb{T}^d \times I$ which is an ordinary instanton.
\\ \\ 
\noindent \textbf{Flat spatial slices.} \\ \\
\noindent \textit{$\mathbb{T}^d$ boundaries, singular solutions.}  We have the solutions
\begin{equation}
	ds^2 = -dt^2 + b^2 |t|^{\frac{4}{d}} \, ds_{\mathbb{T}^d}^2\,,
\end{equation}
for $b >0$ which look like constrained instanton versions of Kasner-type metrics.  There is a big bang/crunch for $t \geq 0$ and $t \leq 0$.
\\ \\ 
\noindent \textbf{Closed spatial slices.} \\ \\
\textit{$\mathbb{S}^d$ boundaries, non-singular solutions.}  Here we have only found an explicit solution in $d = 4$, namely
\begin{equation}
	ds^2 = d\rho^2 + (\rho^2 + b^2) \,ds_{\mathbb{S}^4}^2\,,
\end{equation}
for $b>0$.
\\ \\
\textit{$\mathbb{S}^d$ boundaries, singular solutions.}  We found explicit singular solutions in $d = 2$ and $d = 4$, which are
\begin{align}
	d = 2: \qquad ds^2 &= - dt^2 + b^2 \rho^2 \, ds_{\mathbb{S}^2}^2\,,
	\\
	d = 4: \qquad ds^2 &= -dt^2 + \rho(\rho + b^2) \, ds_{\mathbb{S}^4}^2\,,
\end{align}
with $b>0$.
\\ \\ 
\noindent \textbf{Open spatial slices.} \\ \\
\textit{$\mathbb{H}^d$ boundaries, non-singular solutions.}  An explicit non-singular solution in $d = 4$ is simply
\begin{equation}
	ds^2 = -dt^2 + (t^2 + b^2) \,ds_{\mathbb{H}^4}^2\,,
\end{equation}
for $b>0$.
\\ \\
\textit{$\mathbb{H}^d$ boundaries, singular solutions.} In $d = 2$ and $d = 4$, we find the explicit singular solutions
\begin{align}
	d = 2: \qquad ds^2 &= - dt^2 + b^2 \rho^2 \, ds_{\mathbb{H}^2}^2 \,,
	\\
	d = 4: \qquad ds^2 &= -dt^2 + (\rho^2 + b^2) \, ds_{\mathbb{H}^4}^2\,,
\end{align}
with $b>0$.

\section{Useful formulas for the geometry of the symmetric wormhole with torus boundary}
\label{App:formulas}

Here we collect some useful formulas for differential geometric quantities of the metric corresponding to the symmetric wormhole~\eqref{E:AdSdsymmetric1} with zero twist (and setting $L = 1$). These will be utilized below in Appendix~\ref{App:stability}.  Working in $(\rho, x^1,...,x^d)$ coordinates, let us define
\begin{equation}
	\Delta_{\mu \nu} = 4 \,b^d \,\delta_{\mu \rho} \delta_{\nu \rho} \,\frac{\Lambda}{\sqrt{g}}\,.
\end{equation}
Then for the metric in Eqn.~\eqref{E:AdSdsymmetric1}, we have
\begin{equation}
	\label{E:RicciApp1}
	R_{\mu \nu} = - d \left[\frac{\cosh(d\,\rho)}{1 + \cosh(d\,\rho)}\right] g_{\mu \nu} + \Delta_{\mu \nu}
\end{equation}
and accordingly
\begin{equation}
	\label{E:RicciScalarApp1}
	R = - d (d+1) \left[\frac{\cosh(d\,\rho)}{1 + \cosh(d\,\rho)}\right] + \Lambda \, \text{sech}^{2}\left(\frac{d\, \rho}{2}\right)\,.
\end{equation}
Then we find
\begin{equation}
R - 2 \Lambda = - 2 d \, \left[\frac{\cosh(d\,\rho)}{1 + \cosh(d\,\rho)}\right]
\end{equation}
and
\begin{equation}
	\sqrt{g}\left(R - 2 \Lambda\right) = - 4d\,b^d \cosh(d \,\rho)\,.
\end{equation}
Combining Eqn.'s~\eqref{E:RicciApp1} and~\eqref{E:RicciScalarApp1}, we have
\begin{equation}
R_{\mu \nu} - \frac{1}{2} \, R \, g_{\mu \nu} + \Lambda \, g_{\mu \nu} = \Delta_{\mu \nu}\,.
\end{equation}
Let us also define
\begin{equation}
	\chi_{\alpha \beta \gamma \delta} = \delta_{\alpha \gamma} \delta_{\beta \rho} \delta_{\delta \rho} + \delta_{\beta \delta} \delta_{\alpha \rho} \delta_{\gamma \rho} - \delta_{\alpha \delta} \delta_{\beta \rho} \delta_{\gamma \rho} - \delta_{\beta \gamma} \delta_{\alpha \rho} \delta_{\delta \rho}\,.
\end{equation}
Then we have
\begin{equation}
	R_{\alpha \beta \gamma \delta} = - \tanh^2\left(\frac{d\,\rho}{2}\right) \left( g_{\alpha \gamma} g_{\beta \delta} - g_{\alpha \delta} g_{\beta \gamma}\right) -\frac{d\,b^2}{2^{\frac{d-4}{d}}} \,\text{sech}^{\frac{2(d-2)}{d}}\left(\frac{d\,\rho}{2}\right) \chi_{\alpha \beta \gamma \delta}\,.
\end{equation}

\section{Details of stability analysis}
\label{App:stability}

In this Appendix, we show that the symmetric torus wormhole in Eqn.~\eqref{E:AdSdsymmetric1} is stable against quadratic fluctuations.  In particular, we show that the bulk part of $\frac{\delta^2 S}{\delta g_{\mu \nu} \delta g_{\alpha \beta}}$ is positive-definite.  Further, we establish that minimally coupled free scalar fields, fermions, and gauge fields are also stable.  For our analysis we set $16 \pi G = 1$ and $L = 1$.

Bulk quadratic fluctuations $h_{\mu \nu}$ of the Einstein-Hilbert action in Eqn.~\eqref{E:EHaction1} along with the appropriate constraint term in Eqn.~\eqref{E:constraint1} take the form
\begin{align}
\label{E:qfluct1}
\begin{split}
S_2 = &\int d^{d+1}x \, \sqrt{g}\bigg(- \frac{1}{4} \,\overline{h}^{\mu \nu} \square \overline{h}_{\mu \nu} + \left(\frac{1}{8} - \frac{1}{4(d+1)} \right) h \square h - \frac{1}{2}\left[\nabla^\nu \overline{h}_{\nu \mu} + \left(\frac{1}{d+1} - \frac{1}{2}\right) \nabla_\mu h\right]^2 \\
& \qquad \qquad \qquad \qquad \qquad - \frac{1}{2}\, \overline{h}^{\mu \lambda} \overline{h}^{\nu \sigma} R_{\mu \nu \lambda \sigma} - \frac{1}{2}\left[\overline{h}^{\mu \lambda} \overline{h}_\lambda^\nu - \left(1 - \frac{4}{d+1}\right) h \overline{h}^{\mu \nu}\right]R_{\mu \nu} + \frac{1}{4} \overline{h}^{\mu \nu} \overline{h}_{\mu \nu} (R - 2 \Lambda) \\
& \qquad \qquad \qquad \qquad \qquad \qquad \qquad \qquad \qquad \,\,- \left(\frac{1}{8} - \frac{3}{4(d+1)} + \frac{1}{(d+1)^2}\right) h^2 R - \left(\frac{1}{2(d+1)} - \frac{1}{4}\right) h^2 \Lambda \bigg) \\
&\qquad \qquad \qquad \qquad \qquad \qquad \qquad \qquad \qquad \qquad \qquad \qquad \qquad \qquad \qquad \qquad \,\,\, - i \,\frac{\lambda}{8} \int d^{d+1} x \, \frac{1}{g_{\rho\rho}^{3/2}} \, h_{\rho \rho} h^{\rho \rho}
\end{split}
\end{align}
where
\begin{equation}
h_{\mu \nu} = \overline{h}_{\mu \nu} + \frac{1}{d+1} \, g_{\mu \nu} h
\end{equation}
such that $\overline{h}_{\mu \nu} $ is traceless, and $\square = \nabla_\mu \nabla^\mu$.  Above, we take $\lambda = 8 i \Lambda b^d$.  We have used the notation of~\cite{bastianelli2013one} for convenience.
Let us add a gauge-fixing term to the action, and a compensating ghost term~\cite{bastianelli2013one, christensen1980quantizing}.  The gauge term is
\begin{equation}
	S_{\text{gauge}} = \int d^{d+1} x \, \sqrt{g} \, \frac{1}{2}\left[\nabla^\nu \overline{h}_{\nu \mu} + \left(\frac{1}{d+1} - \frac{1}{2}\right) \nabla_\mu h\right]^2
\end{equation}
which cancels out a term in Eqn.~\eqref{E:qfluct1} above.  The corresponding ghost term is
\begin{align}
	S_{\text{ghost}} &= \int d^{d+1} x \, \sqrt{g} \, c_\mu^* \left( - g^{\mu \nu} \square - R^{\mu \nu} \right) c_\nu 
\end{align}
where $c, c^*$ are ghost fields.

The total action of quadratic fluctuations is $S_{\text{tot}}[h_{\mu \nu}, c_\alpha, c_\beta^*] = S_2[h_{\mu \nu} ] + S_{\text{gauge}}[h_{\mu \nu}] + S_{\text{ghost}}[c_\alpha, c_\beta^*]$. Since the ghosts are Grassmann odd there is no sign constraint on the differential operator appearing in the ghost action, and so it remains to study the quadratic fluctuations coming from $S_2[h_{\mu \nu} ] + S_{\text{gauge}}[h_{\mu \nu}]$, namely
\begin{align}
\label{E:qfluct2}
\begin{split}
&\int d^{d+1}x \, \sqrt{g}\bigg(- \frac{1}{4} \,\overline{h}^{\mu \nu} \square \overline{h}_{\mu \nu} + \left(\frac{1}{8} - \frac{1}{4(d+1)} \right) h \square h - \frac{1}{2}\, \overline{h}^{\mu \lambda} \overline{h}^{\nu \sigma} R_{\mu \nu \lambda \sigma} - \frac{1}{2}\left[\overline{h}^{\mu \lambda} \overline{h}_\lambda^\nu - \left(1 - \frac{4}{d+1}\right) h \overline{h}^{\mu \nu}\right]R_{\mu \nu} \\
& \qquad \qquad \qquad \qquad \qquad \qquad \qquad \quad + \frac{1}{4}\, \overline{h}^{\mu \nu} \overline{h}_{\mu \nu} (R - 2 \Lambda) - \left(\frac{1}{8} - \frac{3}{4(d+1)} + \frac{1}{(d+1)^2}\right) h^2 R - \left(\frac{1}{2(d+1)} - \frac{1}{4}\right) h^2 \Lambda \bigg) \\
&\qquad \qquad \qquad \qquad \qquad \qquad \qquad \qquad \qquad \qquad \qquad \qquad \qquad \qquad \qquad \qquad \qquad \qquad \qquad \,\,\, - i \,\frac{\lambda}{8} \int d^{d+1} x \, \frac{1}{g_{\rho\rho}^{3/2}} \, h_{\rho \rho} h^{\rho \rho}\,.
\end{split}
\end{align}
For $d \geq 1$, the kinetic term $\left(\frac{1}{8} - \frac{1}{4(d+1)}\right)\,h \square h$ is a wrong-sign Gaussian and so it is standard procedure~\cite{gibbons1978path} to deform the contour of integration in the path integral, such as by taking $h \to i \,h$.  However, we need to be careful about doing this in our quadratic action, on account of the constraint term.  In particular, since $h_{\rho \rho}$ is contained in $h$, if our contour deformation were to take $h_{\rho \rho} \to i \,h_{\rho \rho}$ then the term $- i \,\frac{\lambda}{8} \int d^{d+1} x \, \frac{1}{g_{\rho\rho}^{3/2}} \, h_{\rho \rho} h^{\rho \rho}$ would change its overall sign.  This is problematic since upon integrating out $h_{\rho \rho}$ in the path integral, we would get a residual wrong-sign quadratic action in $\lambda$.  But we cannot further rotate the contour of $\lambda$ to rectify this wrong-sign Gaussian since the $\lambda$ integration contour must be parallel to the real axis so as to act as a Lagrange multiplier enforcing the constraint in the original action.  So in summary, we will need to perform some contour deformation involving parts of $h$, but we will not touch the $h_{\rho \rho}$ component to avoid the above issue.

The symmetric torus wormholes possess an $SO(d)$ symmetry, which is broken by the identifications of the boundary tori.  Thus we decompose these metric fluctuations according to their $SO(d)$ symmetry, i.e.~into two scalars $h_{\rho \rho}$ and $h$, the vectors $h_{i \rho}$, and (traceless) tensors $h_{ij} - \frac{1}{d}\,g_{ij} h_{\,\,i}^i$\,.  The two scalars only couple to each other, and the vectors and tensors all individually decouple.

Let us treat the vectors and tensors first.  Writing
\begin{equation}
h_{i\rho}(\rho, \vec{x}) = \sum_{\vec{k}} V_{i,\vec{k}}(\rho) \, e^{i \vec{k} \cdot \vec{x}}\,, \qquad  h_{ij}(\rho, \vec{x}) - \frac{1}{d}\,g_{ij}(\rho, \vec{x}) h_{\,\,i}^i = \sum_{\vec{k}} T_{ij,\vec{k}}(\rho) \, e^{i \vec{k} \cdot \vec{x}}
\end{equation}
with $V_{i,\vec{k}}^*(\rho) = V_{i,-\vec{k}}^*(\rho)$ and $T_{ij,\vec{k}}^*(\rho) = T_{ij,-\vec{k}}^*(\rho)$, the quadratic actions at fixed momentum on the torus are proportional to
\begin{align}
&\int d\rho \left[g^{\frac{d-2}{2d}} \left|V_{i, \vec{k}}'(\rho)\right|^2 + \left(\frac{b^d}{g^{\frac{1}{d}}} \left( 2(2d-3) + 7d\, \cosh(d\rho)\right) + g^{\frac{d-4}{2d}} \vec{k}^2\right) \left|V_{i, \vec{k}}(\rho)\right|^2\right] \\ \nonumber \\
&\int d\rho \left[g^{\frac{d-4}{2d}} \left|T_{ij, \vec{k}}'(\rho)\right|^2 + \left(\frac{b^d}{g^{\frac{2}{d}}} \left( 6 + (7d - 6)\,\cosh(d\rho)\right) + g^{\frac{d-6}{2d}} \vec{k}^2\right) \left|T_{ij, \vec{k}}(\rho)\right|^2\right] 
\end{align}
where \,$'$\, denotes a $\rho$ derivative.  Notice that the above quadratic actions are manifestly positive for $d \geq 2$.

Next we turn to the two coupled scalars.  It is convenient to treat the $d = 2$ and $d > 2$ cases separately.  Let us start with the latter.  Parameterizing
\begin{equation}
h_{\rho \rho}(\rho, \vec{x}) = \sum_{\vec{k}} S_{1,\vec{k}}(\rho) \, e^{i \vec{k}\cdot \vec{x}}\,, \qquad h(\rho, \vec{x}) =  \sum_{\vec{k}} \left(-\frac{2}{d-2} \, S_{1,\vec{k}}(\rho) + 2i \,\sqrt{\frac{d}{d-2}} \, S_{2,\vec{k}}(\rho) \right) \, e^{i \vec{k}\cdot \vec{x}}
\end{equation}
where $S_{1,\vec{k}}^*(\rho) = S_{1,-\vec{k}}(\rho)$ and $S_{2,\vec{k}}^*(\rho) = S_{2,-\vec{k}}(\rho)$, we notice that we have rotated the contour of part of $h$, but have intentionally not rotated the contour for $h_{\rho \rho}$.  We will consider the real part of the corresponding quadratic action, since this is the part relevant for stability.  The resulting quadratic action at fixed momentum on the torus is proportional to
\begin{align}
&\int d\rho \bigg[\frac{d-1}{2(d-2)}  \sqrt{g}\left|S_{1, \vec{k}}'(\rho)\right|^2 + \frac{1}{(d-2)^2}\left(d(d-1)\, b^d\bigg( (d(d-6)+4) + 2(d-2)\, \cosh(d\rho)\right) \nonumber \\
& \qquad \qquad \qquad \qquad \qquad \qquad \qquad \qquad \qquad \qquad \qquad \qquad \qquad \qquad \quad + \frac{1}{2} (d-1)(d-2)\,g^{\frac{d-2}{2d}} \vec{k}^2\bigg) \left|S_{1, \vec{k}}(\rho)\right|^2\bigg] \nonumber \\
+\,\,& \int d\rho \left[\sqrt{g}\left|S_{2,\vec{k}}'(\rho)\right|^2 + \left(b^d \left(\frac{8d}{d-2} + 4 d\cosh(d\rho)\right) + g^{\frac{d-2}{2d}} \vec{k}^2\right) \left|S_{2, \vec{k}}(\rho)\right|^2\right] \,.
\end{align}
Notice that the quadratic action for $S_{1,\vec{k}}$ is manifestly positive for $d > 3$, and the quadratic action for $S_{2,\vec{k}}$ is manifestly positive for $d > 2$.  For the $d = 3$ case, we just need to check the stability of $S_{1,\vec{k}}$ in the $\vec{k} = \vec{0}$ case, since larger $\vec{k}$ will contribute positively to the action and improve stability.  In this case, we find numerically that the minimal eigenvalue of the kernel is $1.61...$ which is indeed positive, and so $S_{1,\vec{k} = \vec{0}}$ and thus $S_{1,\vec{k}}$ for general $\vec{k}$ are stable for $d = 3$.

Finally, we consider the $d = 2$ case for the coupled scalars.  Here we find it convenient to parameterize
\begin{equation}
h_{\rho \rho}(\rho, \vec{x}) = \frac{1}{\sqrt{5}\,b \cosh(\rho)}\sum_{\vec{k}} S_{1,\vec{k}}(\rho) \, e^{i \vec{k}\cdot \vec{x}}\,, \qquad h(\rho, \vec{x}) =  \sum_{\vec{k}} \left(- \frac{1}{\sqrt{5}\,b \cosh(\rho)} S_{1,\vec{k}}(\rho) + \frac{i \sqrt{5}}{b \cosh(\rho)} \, S_{2,\vec{k}}(\rho) \right) \, e^{i \vec{k}\cdot \vec{x}}\,.
\end{equation}
Then the real part of the quadratic action at fixed momentum on the torus is proportional to
\begin{equation}
\int d\rho \,\left[ \left|S_{1,\vec{k}}'(\rho) \right|^2 + \frac{\text{sech}^2(\rho)}{20}\left(-78 + 50 \cosh(2\rho) + \frac{5}{b^2} \, \vec{k}^2\right) |S_{1,\vec{k}}(\rho)|^2 + 10 \,\text{sech}^2(\rho) \left|S_{2,\vec{k}}(\rho)\right|^2\right]
\end{equation}
Notice that $S_{2,\vec{k}}$ has no kinetic term, and has a manifestly positive mass term.  Thus it is quadratically stable.  While the potential term for $S_{1,\vec{k}}$ is not manifestly positive, if we consider the $\vec{k} = \vec{0}$ case numerically we find that the lowest eigenvalue is $0.678...$.  Since non-zero $\vec{k}$ can only improve stability, we conclude that $S_{1,\vec{k}}$ is quadratically stable for all momenta $\vec{k}$.

In summary, we find that the bulk gravitational action is stable to quadratic quantum fluctuations around the symmetric wormhole constrained instanton for $d \geq 2$.  Furthermore, our analysis actually shows that the smallest eigenvalue of $\frac{\delta^2 S}{\delta g_{\mu \nu} \delta g_{\alpha \beta}}$ is positive and bounded away from zero, and so by continuity we have shown that there is a neighborhood in moduli space of quadratically stable constrained instantons containing the symmetric constrained instanton.  We also recall that for $d = 2$, we have previously provided a nonperturbative analysis~\cite{cotler2020ads, cotler2020ads2} which implies stability. 

Having treated the case of pure gravity, we now discuss coupling to matter and gauge fields.  There is no issue of stability for fermions, and the Yang-Mills action is positive semi-definite so there is no likewise issue for gauge fields.  So let us consider the case of a minimally coupled free scalar field.  In this setting, we only need to consider $\rho$-dependent profiles of the field $\phi$, since $x^i$-dependent terms will only contribute to increase the action.  For the case of $\rho$-dependent profiles, the action is manifestly positive semi-definite for $m^2 \geq 0$, so it remains to consider negative $m^2$ greater than or equal to the BF bound.  In this case, the spectrum of the kernel in the quadratic action was analyzed for $d=2$ in~\cite{maldacena2004wormholes}, and it was shown that the eigenvalues were non-negative for $m^2$ above the Breitenlohner-Freedman bound $m_{\rm BF}^2 = -\frac{d^2}{4}$, with a single normalizable zero mode at the BF bound. For configurations of $\phi$ which only depend on $\rho$, the $d > 2$ case follows from the same $d=2$ analysis. The $d>2$ kernel can be mapped to the $d=2$ one by the combination of a rescaling of the radial coordinate $\rho \to \frac{2\rho}{d}$ and of the mass-squared $m^2 \to \frac{4}{d^2}m^2$. Recalling that the BF bound is $m_{\rm BF}^2 = -1$ in $d=2$, we see that these rescalings preserve the BF bound for $d = 2$. The result is that for scalars with $m^2$ above the BF bound in $d\geq 2$, the spectrum of the kernel is non-negative, with a normalizable zero mode appearing when the scalar is at the BF bound.

\section{The instanton action and the boundary stress tensor}
\label{App:On-shell}

In the main text we saw that the holographically renormalized action of some Euclidean wormholes in negative cosmological constant could be simply expressed in terms of the size parameter $b$ and the boundary data. Here we consider any torus wormhole with a metric of the form
\begin{equation}
\label{E:appEmetric}
	ds^2 = d\rho^2 + h_{ij}(\rho)  dx^i dx^j\,,
\end{equation}
and use the modified Einstein's equations~\eqref{E:constrainedEinstein} to derive a simple expression for the renormalized action. That expression is a pure boundary term,
\begin{equation}
\label{E:simpleAction}
	S_{\rm ren} =  \frac{1}{d} \int_{\partial\mathcal{M}} d^dx \sqrt{\gamma} \,\gamma_{ij}T^{ij}\,.
\end{equation}
Here the boundary $\partial\mathcal{M}$ is the union of $\mathcal{B}_1$ and $\mathcal{B}_2$, the two components of the boundary reached as $\rho \to \pm \infty$, and the action is the sum of a boundary term on $\mathcal{B}_1$ and a boundary term on $\mathcal{B}_2$. $\gamma_1$, $\gamma_2$ are the boundary metrics on $\mathcal{B}_1$, $\mathcal{B}_2$, and $T_1$, $T_2$ are the boundary holographic stress tensors given in Eqn.~\eqref{E:holographicStress}. A similar computation shows that the action of wormholes with $\mathbb{S}^1\times \mathbb{S}^{d-1}$ cross-section is, in a natural renormalization scheme, given by~\eqref{E:simpleAction}, although the holographic stress tensor receives a correction relative to~\eqref{E:holographicStress} on account of the boundary curvature.

See Appendix A of~\cite{Stanford:2020wkf} for a similar derivation in a rather general 2D theory of dilaton gravity.

In many contexts the action of a configuration simplifies dramatically when that configuration is on-shell. Our wormholes are not on-shell, yet their action still simplifies enormously. The key fact we will use is that the $ij$ components of the modified Einstein's equations in~\eqref{E:constrainedEinstein} are just the usual $ij$ components of Einstein's equations, which follow from varying the Einstein-Hilbert action with respect to $h_{ij}$ in~\eqref{E:appEmetric}. So the wormhole is partially on-shell, enough to proceed.

The Einstein-Hilbert action may be written in terms of the scalar curvature $R^{(d)}$ of $h$ and the extrinsic curvature $K_{ij}$ of constant-$\rho$ slices as
\begin{equation}
	S_{\rm EH} = -\frac{1}{16 \pi G} \int_{\mathcal{M}} d^d x  d\,\rho \, \sqrt{h} \, \left(R^{(d)} + (\text{Tr}K)^2 - \text{Tr}(K^2) - 2 \Lambda \right) +\frac{1}{8\pi G}\int_{\partial\mathcal{M}}d^dx\, \sqrt{h} \,\text{Tr}(K)\,,
\end{equation}
where
\begin{equation}
	R^{(d)} = 0\,, \qquad K_{ij} = \frac{1}{2} \, \partial_\rho h_{ij}\,, \qquad (\text{Tr} K)^2 =  \frac{1}{4} (h^{ij} \partial_\rho h_{ij})^2\,, \qquad \text{Tr}(K^2) = - \frac{1}{4} (\partial_\rho h^{ij})(\partial_\rho h_{ij})\,.
\end{equation}
We also work in units where $\Lambda = -\frac{d(d-1)}{2}$, setting the AdS radius to unity. Adding the Gibbons-Hawking boundary term $-\frac{1}{8\pi G}\int_{\partial\mathcal{M}}d^dx \sqrt{h}\,\text{Tr}(K)$ we then have
\begin{equation}
	S_{\rm EH} + S_{\rm GH} = S_{\text{bulk}} = -\frac{1}{16 \pi G} \int_{\mathcal{M}} d^d x  d\,\rho \, \sqrt{h} \, \left(  \frac{1}{4} (h^{ij} \partial_\rho h_{ij})^2 + \frac{1}{4} (\partial_\rho h^{ij})(\partial_\rho h_{ij})  - 2 \Lambda \right)\,.
\end{equation}
Let us write the bulk integrand as $\sqrt{h} \, \mathcal{L}$\,.  Varying the action with respect to $h_{ij}$, we find
\begin{equation}
		\delta S_{\rm bulk} = -\frac{1}{16 \pi G} \int_{\mathcal{M}} d^d x  d\,\rho \, \sqrt{h} \left\{ \frac{1}{2}\mathcal{L} \, h^{ij} \delta h_{ij} +\frac{1}{2} (h^{ij} \partial_\rho h_{ij}) \partial_{\rho}(h^{kl} \delta h_{kl}) + \frac{1}{4} (\partial_\rho h^{ij}) \partial_\rho (\delta h_{ij}) + \frac{1}{4} (\partial_\rho h_{ij}) \partial_\rho(\delta h^{ij}) \right\}\,,
\end{equation}
where we have used the usual identity $\delta \sqrt{h} = \frac{1}{2} \, \sqrt{h} \, h^{ij} \delta h_{ij} = - \frac{1}{2} \, \sqrt{h} \, h_{ij} \delta h^{ij}$. Integrating by parts in the second line and performing some cancellations, we have
\begin{align}
\begin{split}
\label{E:deltaSbulk}
	\delta S_{\rm bulk} &= -\frac{1}{16\pi G}\int_{\mathcal{M}}d^dx d\,\rho \left\{ \left(\sqrt{h} \mathcal{L}-\partial_{\rho}(\sqrt{h}h^{kl}\partial_{\rho}h_{kl})\right)h^{ij} - \frac{1}{2}\partial_{\rho}(\sqrt{h}\partial_{\rho}h^{ij})+\frac{1}{2}\partial_{\rho}(\sqrt{h}\partial_{\rho}h_{kl})h^{ik}h^{jl}\right\} \frac{\delta h_{ij}}{2} 
	\\
	& \qquad \qquad + \frac{1}{16\pi G}\int_{\partial\mathcal{M}} d^dx \sqrt{h}\left\{h^{ik}h^{jl}\partial_{\rho}h_{kl}-\left(h^{kl}\partial_{\rho}h_{kl}\right)h^{ij}\right\}\frac{\delta h_{ij}}{2}\,,
\end{split}
\end{align}
so that the $ij$ components of Einstein's equations read
\begin{equation}
	\sqrt{h} \, \mathcal{L} \, h^{ij}  = \partial_\rho\left(\sqrt{h}\, (h^{kl} \partial_\rho h_{kl})\right) h^{ij} +\frac{1}{2} \partial_\rho\left(\sqrt{h} \,\partial_\rho h^{ij}\right)  -\frac{1}{2} \partial_\rho\left(\sqrt{h}\,\partial_\rho h_{kl}\right) h^{ik} h^{j l}\,.
\end{equation}
Contracting both sides with with $h_{ij}$ and dividing by $d$, we find (using $h^{ij}\partial_{\rho}h_{ij} = - h_{ij}\partial_{\rho}h^{ij}$)
\begin{equation}
	\sqrt{h}\,\mathcal{L} =\frac{d-1}{d} \partial_{\rho}\left( \sqrt{h}\,h^{ij}\partial_{\rho}h_{ij}\right) = \frac{2(d-1)}{d} \partial_{\rho}\left(\sqrt{h}\,\text{Tr}(K)\right) \,,
\end{equation}
a total derivative. 

The holographically renormalized action in a setting like this with flat boundary is given by the sum of the Einstein-Hilbert action integrated out to a ``cutoff slice'' near the boundary, along with the Gibbons-Hawking boundary term there and a counterterm proportional to the volume of the cutoff slice.  Then one performs the limit where the cutoff is taken to the conformal boundary. With this limit implicit one writes
\begin{equation}
	S_{\rm ren} = S_{\rm EH} + S_{\rm GH} + S_{\rm CT}\,, \qquad S_{\rm CT} =  \frac{d-1}{8 \pi G}\int_{\partial\mathcal{M}}d^dx \sqrt{h}\,.
\end{equation}
So, adding the counterterm, we see that the instanton action is the pure boundary term 
\begin{equation}
\label{E:renSalmost}
	S_{\rm ren} = -\frac{1}{8\pi G}\frac{d-1}{d} \int_{\partial\mathcal{M}}d^dx \sqrt{h}\,\left( \text{Tr}(K)-d\right)\,.
\end{equation}
This expression can be rewritten in terms of the holographic stress tensor. 

Using the $ij$ equations of motion, the variation of $S_{\rm bulk}$ in~\eqref{E:deltaSbulk} was a pure boundary term, which determines the Brown-York stress tensor
\begin{equation}
	\delta S_{\rm bulk} =\int_{\partial\mathcal{M}} d^dx\sqrt{h}\, T^{ij}_{\rm BY}\frac{\delta h_{ij}}{2}\,, \qquad T^{ij}_{\rm BY} = \frac{1}{8\pi G}\left( K^{ij} - \text{Tr}(K)h^{ij}\right)\,.
\end{equation}
Upon accounting for the volume counterterm, the variation of the full action produces 
\begin{equation}
	\delta S_{\rm ren} = \frac{1}{8\pi G}\int_{\partial\mathcal{M}}\!d^dx\sqrt{h}\left( K^{ij} - \text{Tr}(K)h^{ij} + (d-1)h^{ij}\right)\frac{\delta h_{ij}}{2}\,.
\end{equation}
From this we extract a boundary stress tensor using a defining function. In these coordinates, if there is a conformal boundary at $\rho\to\infty$, then we require a defining function $f\sim e^{\rho}$ to extract a finite boundary metric as
\begin{equation}
	\gamma_{ij} = \lim_{\rho\to\infty} \frac{h_{ij}}{f^2}\,.
\end{equation}
Then we have
\begin{equation}
	\delta S_{\rm ren} = \int_{\partial\mathcal{M}}\!d^dx\sqrt{\gamma} \,T^{ij} \frac{\delta \gamma_{ij}}{2}\,,
\end{equation}
with $T^{ij}$ the holographic stress tensor
\begin{equation}
\label{E:holographicStress}
	\sqrt{\gamma} \,T^{ij} = \frac{1}{8\pi G}\lim_{\rho\to\infty}f(\rho)^{2}\sqrt{h}\left( K^{ij} - \text{Tr}(K)h^{ij}+(d-1)h^{ij}\right)\,.
\end{equation}
It immediately follows that
\begin{equation}
	\frac{1}{d}\sqrt{\gamma}\,\gamma_{ij}T^{ij} = -\lim_{\rho\to\infty}\frac{1}{8\pi G}\frac{d-1}{d}\sqrt{h}\left( \text{Tr}(K)-d\right)\,.
\end{equation} 
Comparing with~\eqref{E:renSalmost} then provides the desired identity~\eqref{E:simpleAction}.

\end{document}